\documentclass[preprint]{article}
\usepackage{spconf,amsmath,graphicx,hyperref}

\title{Application of ASV for Voice Identification after VC and Duration Predictor Improvement in TTS Models}
\begin{document}

\name{%
\begin{tabular}{@{}c@{}}
Borodin Kirill Nikolayevich$^{1}$ \qquad
 Kudryavtsev Vasiliy Dmitrievich$^{1}$ \\ Mkrtchian Grach Maratovich$^{1}$
\qquad Gorodnichev Mikhail Genadievich$^{1}$\\ Korzh Dmitrii Sergeevich$^{2,3}$
\end{tabular}}

\address{
  $^1$ Moscow Technical University of Communication and  Informatics, Moscow, Russia\\
  $^2$ Artificial Intelligence Research Institute, Moscow, Russia \\
  $^3$ Skoltech, Moscow, Russia
 }

%
%
%
\ninept
\maketitle
\begin{abstract}

\par One of the most crucial components in the field of biometric security is the automatic speaker verification system, which is based on the speaker's voice. It is possible to utilise ASVs in isolation or in conjunction with other AI models. In the contemporary era, the quality and quantity of neural networks are increasing exponentially. Concurrently, there is a growing number of systems that aim to manipulate data through the use of voice conversion and text-to-speech models. The field of voice biometrics forgery is aided by a number of challenges, including SSTC, ASVSpoof, and SingFake.

\par This paper presents a system for automatic speaker verification. The primary objective of our model is the extraction of embeddings from the target speaker's audio in order to obtain information about important characteristics of his voice, such as pitch, energy, and the duration of phonemes. This information is used in our multivoice TTS pipeline, which is currently under development. However, this model was employed within the SSTC challenge to verify users whose voice had undergone voice conversion, where it demonstrated an EER of 20.669.

\end{abstract}
\section{Introduction}\label{sec:intro}
\par Voice conversion is a technique employed to alter the voice of one individual to that of another. VC models are employed in the enhancement of TTS systems, the alteration of the voice of call centre operators with the objective of increasing sales, and the transformation of the human voice into that of fictional characters. However, these models can also be employed for unscrupulous purposes, such as deceiving bank employees, perpetrating fraud by impersonating relatives, and committing identity forgery. 

\par The process of text-to-speech conversion involves the transformation of text into audio that is intended to resemble human speech. A high-quality TTS model is capable of producing a voice that is genuinely real-life in its likeness, and thus indistinguishable from the real thing.Such systems are employed in a variety of fields, enabling the automation of processes that were previously infeasible. Examples include the voice-acting of books and advertisements, websites and applications for visually impaired individuals, and so forth. However, as with VC models, such architectures can also be employed for the purpose of deception.

\par The advent of modern technology has led to the rapid development of systems capable of detecting modified and synthetic voice. Such models are referred to as 'antispoofing' models.  The community ASVspoof(\cite{EURECOM+4573}, \cite{delgado18_odyssey}, \cite{wang2020asvspoof}, \cite{yamagishi2021asvspoof}), which regularly organises challenges and collects data corpuses to study the detection of voice biometric forgery, plays an active role in the development of such systems. It is also important to note the contribution of the SingFake competition\cite{zang2024singfake}, which highlighted the significant shortcomings of SOTA models on the ASVSpoof benchmark. In particular, the researchers identified two key issues with the models in question. Firstly, they noted that the models lack generalisability in the context of music. Secondly, they observed that the models are not as effective when applied to different languages. This illustrates the need for further research and development of models that can effectively detect spoofing in voice biometrics.

\par The detection of voice spoofing is of great import, as effective detection can prevent a multitude of malicious activities involving the use of a spoofed voice. However, the problem of identifying the attacker is not insignificant. Thus, if the original voice can be identified in a fake voice that may have been obtained using a voice conversion system, a scammer can be identified. 

\par The current state of research in this area is insufficient. In this context, the SSTC challenge is of particular importance, as it provided researchers with the opportunity to explore in depth methods for identifying the original voice in fake audio recordings. The database facilitates research in this field, thereby facilitating the emergence of novel methods for detecting the original voice of an attacker, which in turn enhances the overall level of security in the field of voice biometrics.

\par In order to construct a TTS system, it was deemed necessary to obtain a number of crucial speech characteristics of the speaker. In order to achieve this, a system for automatic speaker verification was employed, with the use of special encoders to obtain as much useful speaker information as possible. This paper describes a model that discriminates speakers acceptably and its embeddings contain sufficient information to predict various voice characteristics. Furthermore, the impact of the model's embeddings on predicting the duration of each phoneme is also described. In addition, we participated in an SSTC challenge in which we were required to verify voices that had undergone voice conversion. The outcomes of this Challenge, along with a comprehensive overview of the system, are presented in this paper.

\section{Model encoders}
\label{sec:encoders}

\par A series of encoders, each configured to accept a different audio frontend, were employed to extract high-level feature maps. Sound representations were employed in this study, including the Constant-Q Transform (CQT), Mel spectrogram, and Pitch spectrogram. This section provides a detailed description of the implementation of models for the primary features from the previously mentioned audio transformations.

\subsection{CQT Encoder}\label{ssec:cqt}

\par The Non-stationary Gabor Transform algorithm \cite{cqt}, with the following parameters, was employed to obtain the CQTs:
\begin{itemize}
    \item \textit{Number of octaves = 8}
    \item \textit{Number of bins per octave = 64}
    \item \textit{Block length = 1 second}
\end{itemize}

\par The obtained sound representation was encoded using a 10-block Specblock encoder. Each Specblock comprises a complex convolution layer, a complex ELU activation function and a complex dropout.
\par Fixed configuration of complex convolutional layers:
\begin{itemize}
    \item \textit{kernel size = 3}
    \item \textit{stride = 2}
    \item \textit{padding = 1}
\end{itemize}

\par Complex ELU is an activation function for complex numbers that is applied by separately applying ELU to the real and imaginary parts. The resulting components are then recombined into a single complex number. The complex dropout operates in a manner analogous to that of a regular dropout: a random selection process, with a probability of 0.4, results in the nullification of some parameters.

\par Following the encoding process, the real and imaginary parts are concatenated along the first dimension. The configuration of the increase in the number of channels in the convolutional layers is presented in Table. \ref{tab:channels}.

\begin{table}[h]
    \centering
    \begin{tabular}{ccc}
        \hline
        block number & input channels & output channels \\
        \hline
        1 &1  &  32\\
        2 & 32 &  32\\
        3 & 32 &  32\\
        4 & 32 & 32 \\
        5 &32  &  64\\
        6 & 64 & 64 \\
        7 & 64 & 128 \\
        8 & 128 & 128 \\
        9 & 128 & 128\\
        10 & 128 & 128 \\

        \hline
    \end{tabular}
    \caption{Channel configuration}
    \label{tab:channels}
\end{table}

\subsection{Mel-spectrogram Encoder}\label{ssec:mel}
\par To obtain the mel spectrogram, we employed the following parameters for conversion:
\begin{itemize}
    \item \textit{Size of Fast Fourie Transform, as well as the window size = 1201}
    \item \textit{Hop length = 600}
    \item \textit{Number of mel filterbanks = 128}
    \item \textit{Type of window - Hann}
\end{itemize}

\par In order to obtain high-level feature maps from mel-spectrograms, a Vision Transform(ViT) \cite{dosovitskiy2021image} with the following configuration was employed:
\begin{itemize}
    \item \textit{Embedding dimensionality = 256}
    \item \textit{Hidden dimensionality = 512}
    \item \textit{Number of attention heads = 32}
    \item \textit{Number of layers = 3}
    \item \textit{Output dimensionality = 512}
    \item \textit{Patch size = 16}
    \item \textit{Number of patches = 320}
\end{itemize}

\subsection{Pitch Encoder}\label{ssec:pitch}
\par The pitch spectrogram was obtained by first utilising the distributed inline filtering with overlap (DIO) algorithm to generate a series of continuous pitch contours. Subsequently, the Contious Wavelet Transform(CWT)(\cite{pitch}, \cite{Hirose2015SpeechPI}, \cite{Lee2019}) was applied. The measurements were conducted in scale factors within the range [1, 37]. The resulting pitch spectrogram was run through the VIT with the following parameters:
\begin{itemize}
    \item \textit{Embedding dimensionality = 256}
    \item \textit{Hidden dimensionality = 512}
    \item \textit{Number of attention heads = 32}
    \item \textit{Number of layers = 3}
    \item \textit{Output dimensionality = 512}
    \item \textit{Patch size = 9}
    \item \textit{Number of patches = 356}
\end{itemize}

\section{model and loss function}
\label{sec:model}

\par The input data for the model is a 4-second audio segment. To ensure the uniformity of input data, the data were resampled to 24,000 before being fed into the model. The segment of length 4 was selected at random from the original audio recording.

\par The prepared audio is fed to the Pitch encoder, the CQT encoder and the mel-spectrogram encoder. The resultant vectors are then concatenated along first dimension following their passage through the encoders. They are then passed through a fully-connected layer with an output feature count of 2048. The results obtained are passed through the ELU activation function. This process involves the extraction of primary embeddings.

\par The model was trained using the Additive Margin Softmax (AM-Softmax) \cite{Wang_2018} loss function with a parameter scale of 30 and a margin of 0.4. The loss function is presented in eq. \ref{loss:am}

\begin{equation}\label{loss:am}
    \mathcal{L} = - \frac{1}{n} \sum^n_{i=1} \log \frac{e^{30(W^T_{y_i}f_i - 0.4)} }{e^{30(W^T_{y_i}f_i - 0.4)} + \sum^c_{j=1, j\neq y_i} e^{30W^T_j f_i}}
\end{equation}

\section{EXPERIMENTS AND RESULTS OF THE MODEL AS PART OF TTS}
\label{sec:tts_results}
\subsection{Datasets description}\label{ssec:dataset_tts}

\par To test the model, we used an open dataset\footnote{https://www.kaggle.com/datasets/kongaevans/speaker-recognition-dataset} containing the speeches of the following five speakers: Benjamin Netanyahu, Jens Stolnberg, Julia Gillard, Margaret Tarcher and Nelson Mandela. There are 1,500 speech samples for each speaker. The dataset was collected using open data from American Rhetoric\footnote{https://www.americanrhetoric.com/speechbank.htm } (online speech bank).

\par Despite the large number of samples used, using only 5 to test ASV is not sufficient. In addition to the above dataset, we also used the CMU ARCTIC\cite{kominek04b_ssw}, which consists of 18 speakers, each of which has 500-1000 speech samples collected.

\par We have used LibriTTS-R \cite{koizumi2023librittsr} for the training of the model. From the train subset of this dataset, 800 speakers were randomly selected, with 200 utterances selected for each speaker. 

\subsection{Metrics description}\label{ssec:metrics_tts}

\par Since the model was trained to ensure that the embeddings of one person were close to each other in the multidimensional space and those of different people were far apart, we needed to use a metric that would estimate the ratio of cosine similarities between identical pairs of voices and between different pairs of voices at a certain threshold. 

\par We chose TPR @ FPR. The idea behind this metric is that, given a fixed rate of false positive pairs, we count how many true positive pairs we have before this last allowed false positive pair, and then divide the resulting number by the total number of true positive pairs. The algorithm used to calculate this metric:
\begin{enumerate}
    \item Fixing the threshold value TPR @ FPR = N.
    \item Calculate the cosine similarity between all positive pairs.
    \item Calculate the cosine similarity between all negative pairs.
    \item Calculate how many elements make up N times the number of false positive pairs. This number is k.
    \item Sort all cosine similarity values of false pairs in descending order. The K-th number is the threshold distance.
    \item Count the number of positive pairs whose cosine distance is less than the threshold distance. 
    \item Divide the number obtained by the total number of positive pairs.
\end{enumerate}

\subsection{Loss function selection}\label{ssec:am_vs_arc}

\par The choice was between ArcFace \cite{Deng_2022} with scale=30, margin=0.5 and AM-Softmax \ref{sec:model}. The model was trained and tested on non-overlapping subsets of the test dataset for 5 epochs. The result is shown in Table \ref{tab:losses}

\begin{table}[h]
    \centering
    \begin{tabular}{ccc}
        \hline 
        metric/loss   & ArcFace & AM-Softmax         \\
        \hline
        TPR@FRP=0.5   &  0.7134 &  \textbf{0.906}   \\
        TPR@FRP=0.2   &  0.3785 &  \textbf{0.6846}  \\
        TPR@FRP=0.1   &  0.2429 & \textbf{0.5101}   \\
        TPR@FRP=0.05  &  0.1560 &  \textbf{0.3960}  \\
        TPR@FRP=0.01  &  0.04291& \textbf{0.2148}   \\
        \hline
    \end{tabular}
    \caption{Different loss function training results}
    \label{tab:losses}
\end{table}

\par Based on these results, we decided to use AM-Softmax.

\subsection{Training results}\label{ssec:results_tts}

\par We trained the model for 30 epochs, for a total training time of 30 hours. The training was carried out using x1 Tesla A100 40GB. The checkpoint model at the epoch when it exhibited the highest TPR@FPR=0.01 on the validation subsample was selected for evaluation. The outcomes of the model on the test set are presented in Table \ref{tab:training}.

\begin{table}[h]
    \centering
    \begin{tabular}{cc}
        \hline 
        metric   & value \\
        \hline
        TPR@FRP=0.5   &  0.9869 \\
        TPR@FRP=0.2   &  0.9469 \\
        TPR@FRP=0.1   &  0.9044 \\
        TPR@FRP=0.05  &  0.8354 \\
        TPR@FRP=0.01  &  0.5920 \\
        \hline
    \end{tabular}
    \caption{Results on testing subsample}
    \label{tab:training}
\end{table}

\par
\begin{figure}[h]
    \centering
    \includegraphics[, clip,width=\linewidth]{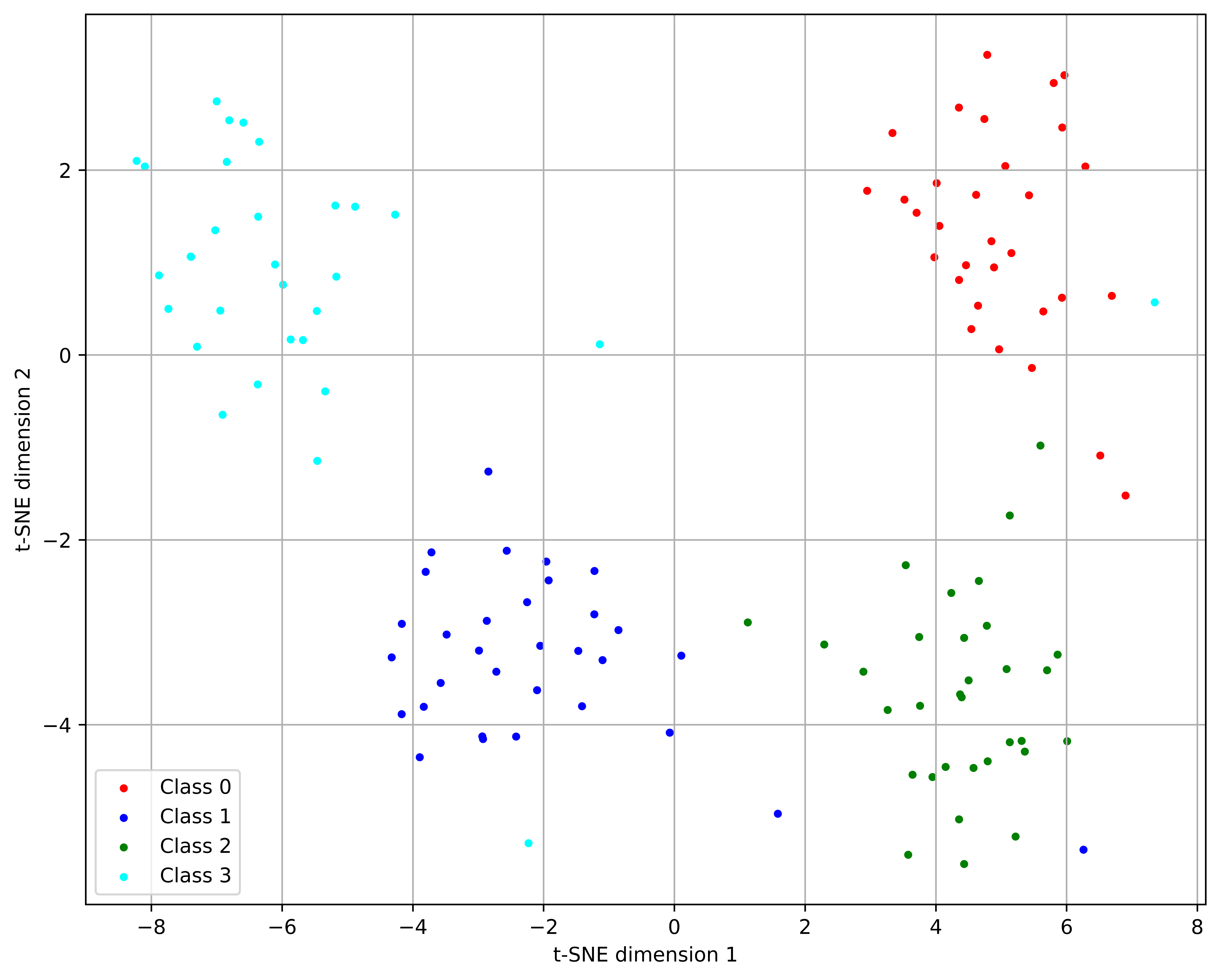}
    \caption{Embeddings distribution}
    \label{fig:tsne}
\end{figure}

 Furthermore, we proceeded to test our model on in-the-wild data. Four distinct voices were recorded from open sources, ensuring that they were not included in any of the datasets used in the experiments. The voices were divided into two categories, male and female, in a one-to-one ratio. A two-minute audio segment was divided into a four-second fragment with no overlap. The distribution in the embedding space obtained with t-SNE is illustrated in Figure \ref{fig:tsne}.

\subsection{Impact of ASV embeddings on the duration predictor}\label{ssec:duration_predictor}

\par The duration predictor is one of the key components of our TTS Pipeline, which is currently under development. The objective of the duration predictor is to predict the duration of pronunciation of phonemes by a specific speaker. As this work does not extend to the deepening of TTS, for the sake of convenience, we refer to the duration predictor architecture as a black box. The black box architecture is immutable and can be trained.

\par The target values were obtained from the annotation generated by the Montreal Forced Aligner \cite{mfa}. For each phoneme, the length of the corresponding mel-spectrogram frame was calculated using the following transformation parameters:
\begin{itemize}
    \item \textit{Size of Fast Fourie Transform, as well as the window size = 1024}
    \item \textit{Hop length = 256}
    \item \textit{Number of mel filterbanks = 128}
    \item \textit{Type of window - Hann}
\end{itemize}

\par Four experiments were conducted in order to investigate the effect of embeddings on black box predictions. The following experiments were conducted, with the type of input differing as follows: encoded phonemes; encoded phonemes concatenated with a random noise of embedding size; encoded phonemes concatenated with an embedding of the same speaker; encoded phonemes concatenated with an embedding of the same utterance.

\par The following metrics were employed for the evaluation of the models: MAE, RMSE, weighted duration error rare and the concordance correlation coefficient \cite{ccc}. The weighted DER metric was devised to quantify a range of model errors. Given that phoneme duration is a relatively subjective value, with each individual pronouncing the same word in a distinct manner, we have devised a metric that allows for minor deviations from the target values to be compensated for, while larger discrepancies are reflected in a larger error coefficient, as outlined in eq. \ref{eq:f_wder}, \ref{eq:wder}

\begin{equation} \label{eq:f_wder}
    f_{wder}(x, \hat{x}) = \begin{cases}
    1 , &|{x=\hat{x}}| \leq 1 \\
    |x - \hat{x}|^{-1},& otherwise
    \end{cases}
\end{equation}

\begin{equation} \label{eq:wder}
    WDER = \frac{1}{N} \sum^n_{i=0} f_{wder}(x_i, \hat{x_i})
\end{equation}

\par The results of the model following four experiments are presented in Table \ref{tab:dp}.

\begin{table}[h]
    \centering
    \begin{tabular}{ccccc}
        \hline
        experiment/result                      & MAE $\downarrow$ & RMSE$\downarrow$  & WDER$\uparrow$   & CCC$\uparrow$   \\
        \hline
        encoded phonemes                       & 2.045            & 3.836             & 0.9309           & 0.771           \\\\
        \shortstack{encoded phonemes\\+random noise}        & 2.198            & 4.366             & 0.9364           & 0.7328          \\\\
        \shortstack{encoded phonemes\\+speaker embedding}   & 1.87             & 3.607             & 0.949            & 0.801           \\\\
        \shortstack{encoded phonemes\\ +utterance embedding} & \textbf{1.869}   & \textbf{3.503}    & \textbf{0.9501} & \textbf{0.8038} \\
        \hline
    \end{tabular}
    \caption{The results of the duration predictor experiments}
    \label{tab:dp}
\end{table}

\par As illustrated in Table \ref{tab:dp}, the results demonstrated that the embeddings of the same utterance exhibited the most optimal performance, indicating that the phoneme length information is indeed retained within the embeddings. Concurrently, the results with embeddings of another utterance by the same speaker also demonstrated good results, indicating the retention of phoneme duration information. The experiment in which random noise was added to the embeddings demonstrated a decline in the results, which lends support to the hypothesis that it is the speaker's embeddings that retain phoneme length information. 

\section{EXPERIMENTS AND TRAINING RESULTS FOR SSTC}
\label{sec:sstc_results}
\subsection{Dataset description}\label{ssec:dataset_sstc}

\par The SSTC dataset utilises the Librispeech \cite{libri} train-clean-100, train-clean-360, dev.clean, test.clean datasets as the source speaker dataset, and the VoxCeleb 2 dev \cite{Chung_2018}, VoxCeleb 1 test \cite{Nagrani_2017} datasets as the target speaker datasets. 

\par The dataset was designed to identify the original speaker in modified audio. The audio was modified using 16 different voice conversion models.

\subsection{Model training}\label{ssec:sstc_training}

\par A detailed description of the architectural design of the model used in this Challenge can be found in Section \ref{sec:model}. The model was trained exclusively on the LibriSpeech corpus. It should be noted that no pre-training was applied to the data from other sources. The model was trained for 30 epochs on a Tesla A100 with 40 GB.

\par The model was trained on the SSTC training subset for 22000  steps, one epoch lasts 9100 steps, or 22 hours 34 minutes. A batch size of 52 items was employed, with the number of classes for AM-Softmax being 1172. For the purposes of training, the x3 Tesla A100 40 GB was employed, for validation and test x1 Tesla P100 12Gb. Due to the limited timeframe of the Challenge, it was not possible to conduct training on more epochs. 
The results of the training are presented in Table \ref{tab:sstc}.

\begin{table}[h]
    \centering
    \begin{tabular}{cc}
        \hline 
        subset   & EER \% \\
        \hline 
        development   &  46.981 \\
        test   &  44.809 \\
        \hline
    \end{tabular}
    \caption{Results on SSTC}
    \label{tab:sstc}
\end{table}

\subsection{Model ensembling}\label{ssec:ensembling}

\par To enhance the performance of our model, we decided to apply the stacking ensemble technique. As a second algorithm for model composition, we selected the baseline model, which was provided by the organisers of the competition. As a solver algorithm, we employ a variety of averaging techniques. The final results, denoted by $Y_i$, were obtained by combining the vector of results from our model, $X_1$, with those from the second model, $X_2$. This was done in accordance with the following eq. \ref{eq:y1}, \ref{eq:y2}, \ref{eq:y3}:

\begin{equation}\label{eq:y1}
    Y_1 = \frac{1}{2} (X_1 - 0.99) * 100 + \frac{1}{2} X_2
\end{equation}

\begin{equation}\label{eq:y2}
    Y_2 = 0.35 * (X_1 - 0.99) * 100 + 0.65 * X_2
\end{equation}

\begin{equation}\label{eq:y3}
    Y_3 = 0.25 * (X_1 - 0.99) * 100 + 0.75 * X_2
\end{equation}

The results for these ensemble techniques are presented in Table \ref{tab:results}.

\begin{table}[h]
    \centering
    \begin{tabular}{cc}
        \hline 
        ensemble technique   & EER \% \\
        \hline 
        $Y_1$   &  23.625 \\
        $Y_2$   &  20.762 \\
        $Y_3$   &  20.669 \\
        \hline
    \end{tabular}
    \caption{Results of different ensemble techniques.}
    \label{tab:results}
\end{table}

\par We found that stacking two models allowed us to improve the quality of predictions on the test sample by almost two times.

\section{conclusion}
\label{sec:conclusion}

\par In this work, we explored the potential of an automatic speaker verification system for a range of applications. Our findings suggest that the model trained for user verification may offer insights that could enhance the performance of text-to-speech models in terms of duration prediction. 

\par We also explored the possibility of training the model to identify the original speaker in audio that has been modified with voice conversion. While we were able to achieve an EER=20, we believe that the dataset proposed in the SSTC competition offers a promising field for further research.

\clearpage
\bibliographystyle{IEEEbib}
\bibliography{main}

\end{document}